# Surface modification of fly ash by mechano-chemical treatment

Kunihiko Kato [a], Yunzi Xin [a], Takashi Hitomi [b], Takashi Shirai [a*]

[a] *Advanced Ceramics Research Center, Nagoya Institute of Technology, Gokiso, Showa-ku, Nagoya, Aichi 466-8555 Japan*
[a] *Construction System and Materials Department, Technical Research Institute, Technology Division, Obayashi Corporation, 640, Shimokiyoto 4-Chome, Kiyose-shi, Tokyo 204-8558 Japan*
*corresponding author E-mail address: shirai@nitech.ac.jp*

## Abstract

Fly ash (FA), as an industry by-product has attracted much attention as a suitable supplier of silicon (Si) and aluminum (Al) in preparation of geopolymer for the sustainable environment and material science applications. Here, the effect of mechano-chemical (MC) treatment for surface modification of FA powder was systemically investigated by analyzing the size, surface morphology, crystal structure and dissolubility of Si and Al ions in alkali solutions. The dissolution dynamic as a function of MC treatment time is discussed in details and concluded with a recombination model of "grinding effect" and "activation effect", which can be well correlated with the change of surface morphology and crystal structure, respectively.

# 1 Introduction

Fly ash (FA), as an industrial byproduct from thermal power plant, is worldwide produced on a large scale. The excess landfilled of FA in ash disposal sites may cause serious space problems, potential risks of air and water pollution due to leaching [1-6]. Consequently, the efficient usage of FA has attracted much attention with respect to the sustainable environment and material science. It has been reported that FA can be used as suitable silicon and alumina supplier in the preparation of geopolymer [7-9], because quartz and mullite components are mainly contained in FA [10-12]. However, surface modification and activation are necessary for enhancing the dissolubility in highly concentrated alkali solution (8-14 M) [13-16] due to the relatively poor solubility of the FA raw powder.

Mechano-chemical (MC) treatment via milling process is well known as a key technology to modify the surface and improve reactivity of FA in alkali solution. It has been previously reported that the mechanical strength of geopolymer dramatically increases by utilizing FA modified by MC treatment [17-20]. However, the correlation between dissolubility of MC treated FA in alkali solution and the change of morphology and crystal structure of FA has not been clarified yet. Herein, the surface of FA is modified by MC treatment and the size, the specific surface area and crystal structure the dissolubility of MC treated FA in alkali solution were investigated, whose results were well correlated by modeling the dissolution dynamic as recombination of "grinding effect" and "activation effect". The results in present study provide important information for modifying and designing of FA surface during MC treatment.

## 2 Experimental procedures

### 2.1　MC treatment and characterization of FA powder

FA powder (J-powder, JPEC, Japan) was used as raw material. MC treatment was performed on a commercial planetary ball mill (Pulverisette-5, Fritsch, Germany) using zirconia ball (Φ=10mm) and pot. MC treated FA powders were prepared under the following conditions: revolution rate of 300 rpm; milling time of 1, 6 and 24 h (which were identified as MC-1, MC-6 and MC-24, respectively.).

Particles morphology of FA powder were observed by FE-SEM (JSM-7000F, JEOL, Japan). The particle size distribution in solution was measured by a laser diffraction particle size analyzer (Microtrac MT3200IIseries, NIKKISO, Japan). The specific surface area was analyzed by measuring the $N_2$ adsorption / desorption isotherms on high precision gas adsorption measurement instrument (BELSORP-max2, MicrotracBEL, Japan) at 196°C (77 K). The crystal structure was identified from the pattern measured by X-ray diffraction instrument (XRD; Model RINT 1000, Rigaku, Japan). A Cu Kα line with λ= 1.5406 Å was used with operating voltage of 40 kV and current of 40 mA.

**2.2 Evaluation of dissolubility of FA raw and MC treated powder in alkali solution**

The dissolution amount of cation $Si^{4+}$ and $Al^{3+}$ of FA powder in alkaline solution were measured by inductivity coupled plasma optical emission spectrometer (ICP-OES: SPS-7800, SII Nano Technology, Japan). The sample for ICP analysis was prepared by the following condition. Firstly, 0.1 g of powder was dissolved in 10 g of 3 mol/l NaOH solution (ikkyu, wako, Japan), followed by shaking at room temperature (25°C) for 48 h (Shaking baths SB20, AS ONE, Japan). Then the undissolved FA were removed by centrifugation at 15000 rpm for 20 min (KOKUSAN type, Japan). The obtained supernantant was filtered through a 0.2 μm membrane filter to remove trace remaining particles and the solution was diluted 10 time with distilled water.

# 3 Results and discussion
## 3.1 Characterization for FA powder after MC treatment

The SEM images of raw and MC treated FA powders are shown as Figure 1, which demonstrates that the morphology after MC treatment is significantly changed and particle size decreased as milling time increased. The shape of particles changed from spherical to non-spherical due to grinding and agglomeration (inset in Figure 1 (a)-(d)). With further measurement of particle size distributions (Figure 2), the mean particle size ($d_{50}$) of raw, MC-1, MC-6 and MC-24 were 9.67, 2.00, 1.92 and 3.51 μm, respectively. MC treatment significantly decreased the particle size in all conditions, while such size was increased again upon longer MC time (as MC-24). These results show good agreement with SEM images, where we also observed 1μm fine particle in huge secondary particles of 30-40 μm in Figure 1 (d). The specific surface areas (SSA) of samples are displayed in Figure 3. The SSA of FA powders increased after MC treatment (Figure 3), whereas the value of SSA did not linearly correlated with milling time due to the agglomeration among particles. Figure 4 shows XRD patterns of raw and MC treated FA powders. Mullite and Quartz crystal phases were observed in all FA powders. As MC treatment time increased, the intensity of both crystal peaks was reduced.

**3.2 Evaluation of solubility of FA powder in alkali solution with MC treatment**

Figure 5 (a) shows the dissolution amount of $Al^{3+}$ and $Si^{4+}$ from FA powders in 3M NaOH solution. As a result, the amount of both $Al^{3+}$ and $Si^{4+}$ ions are drastically increased upon MC treatment. Furthermore, the amount per unit surface area is also summarized as Figure 5 (b), which indicates that the amounts of dissolute ions were increased with longer MC treatment (Figure 5 (b)). Here, the dissolution dynamic can be considered as a recombination model of "grinding effect" and "activation effect". The change of surface morphology can be attributed as "grinding effect", and the change of crystallinity can be due to "activation effect". The morphology and crystallinity change can be evaluated by comparing the SSA and degree of amorphous component, respectively. Based on the above dissolution dynamic, the increase of dissolution amount by MC treatment should be expressed by equation (1).

$$\Delta n = \Delta n_g + \Delta n_a \qquad (1)$$

Here, Δn is the total increased dissolution amount upon MC treatment, $\Delta n_g$ and $\Delta n_a$ are the increased amount caused by "grinding effect" and "activation effect", respectively. When $\Delta n_g$ is proportional to the increase rate of SSA from S0, $\Delta n_g$ can be expressed by equation (2).

$$\Delta n_g = n_{raw} \times \left( \frac{SSA_{MC}}{SSA_{raw}} - 1 \right) \qquad (2)$$

As for equation (2), $\Delta n_g=0$ when the SSA of MC treated FA does not change from raw powder, which demonstrates that the "grinding effect" is negligible. Δn can also be expressed by equation (3).

$$\Delta n = n_{MC} - n_{raw} \qquad (3)$$

Therefore, $\Delta n_a$ can be derived as equation (4) by using the equations from (1) to (3).

$$\Delta n_a = \Delta n - \Delta n_g = n_{MC} - n_{raw} - n_{raw} \times \left( \frac{SSA_{MC}}{SSA_{raw}} - 1 \right)$$

$$= n_{MC} - n_{raw} \times \frac{SSA_{MC}}{SSA_{raw}} \qquad (4)$$

Figure 6 (a) and (b) show the value of $\Delta n_g$ and $\Delta n_a$ for $Al^{3+}$ and $Si^{4+}$ ion, respectively. It can be concluded that "grinding effect" is the dominant factor for the increased dissolution amount in the case of short MC treatment. On the other hand, "activation effect" became dominant during long MC treatment, such as 6 and 24 hours. These phenomena can be interpreted as in Figure 7 and following three stages: (1) at the initial stage of MC treatment, the mechanical energy was used to grind coarse particles; (2) at the middle stage, the mechanical energy was not only used for grinding of particles but also cause the amorphization of crystal when coarse particles disappeared and reached around grinding limit; (3) at the final stage, the energy would be applied for grinding the coarse secondary particle, since fine particle formed aggregates and/or agglomerate by strong impact from mechanical energy. As described above, it can be concluded that it is very important to optimize the MC treatment conditions in order to achieve proper surface modification of particles by considering the recombination of grinding and activation effect.

As an additional experiment, the dissolubility of FA powders (raw and MC-24) in NaOH solution was investigated for different concentrations (0.05, 0.1, 0.5, 1, 3, 8 M) (Figure 8). It clarified that the dissolution amount of $Al^{3+}$ and $Si^{4+}$ MC treated powder of MC-24 in NaOH solution were comparable to raw powder, whose difference was significantly increased in the case of NaOH solution with higher concentration, such as 8 M. Therefore, MC treatment drastically improved the dissolubility of FA powder.

# 4 Conclusions

In present study, the effect of MC treatment on surface modification of FA powder were systemically investigated by analyzing size, surface morphology, and crystal structure of the dissolution amount of $Si^{4+}$ and $Al^{3+}$ in alkali solutions. Upon MC treatment, the size of FA particles was decreased while the size of secondary particle increased as agglomeration occurred. The dissolution amount $Si^{4+}$ and $Al^{3+}$ ions were increased upon MC treatment. The dissolution dynamic as a function of milling time was considered as recombination model of "grinding effect" and "activation effect", which can be attributed to the change of surface morphology and crystal structure, respectively. According to the calculated dissolution amount for "grinding effect" and "activation effect" respectively, the MC treatment can be considered as initial, middle and final stages according to the different role of mechanical energy applying for grinding of particle or activation of particle surface.

# Figure captions

Figure 1. The particle morphology of raw and MC treated powders (MC-1-MC-24) (in FE-SEM): (a) raw, (b) MC-1, (c) MC-6 and (d) MC-24. Inset is SEM images in high magnification.

Figure 2. The particle size distribution of raw and MC treated powders (MC-1-MC-24). Inset is the mean particle size ($d_{50}$).

Figure 3. The specific surface area of raw and MC treated powders (MC-1-MC-24) calculated by BET methods from $N_2$ adsorption/desorption isotherms.

Figure 4. The XRD patterns of raw and MC treated powders (MC-1-MC-24).

Figure 5. The dissolution amounts of ions into 3M NaOH solution (ICP-OES): (a) The dissolution amount of $Si^{4+}$ and $Al^{3+}$, (b) The dissolution amount of $Si^{4+}$ and $Al^{3+}$ per unit surface area.

Figure 6. The dependence of dissolution amount on effects of milling and activation: The dissolution amount of (a) $Al^{3+}$ and (b) $Si^{4+}$.

Figure 7. The schematic illustration of particle behavior during MC treatment.

Figure 8. The dissolution amount with various concentration of NaOH solution: The dissolution amount of (a) $Al^{3+}$ and (b) $Si^{4+}$.

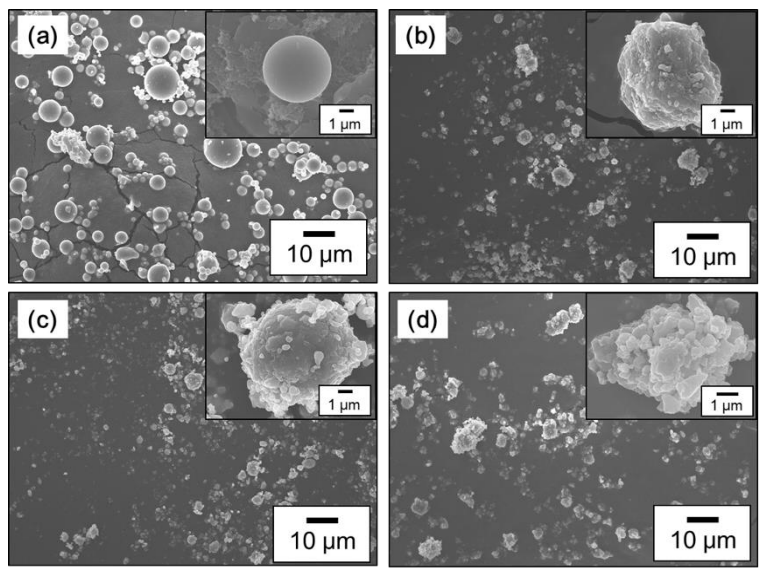

Figure 1. The particle morphology of raw and MC treated powders (MC-1-MC-24) (in FE-SEM): (a) raw, (b) MC-1, (c) MC-6 and (d) MC-24. Inset is SEM images in high magnification.

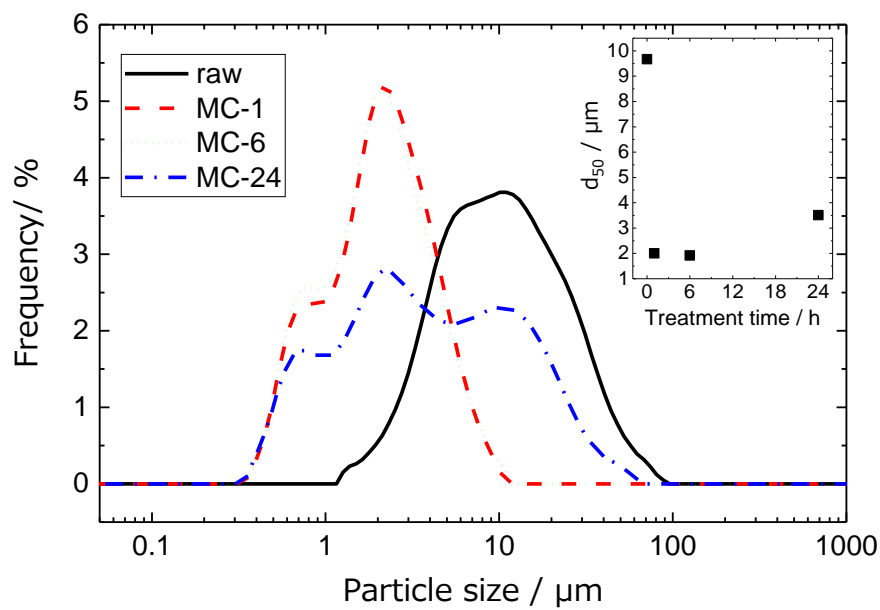

Figure 2. The particle size distribution of raw and MC treated powders (MC-1-MC-24). Inset is the mean particle size ($d_{50}$).

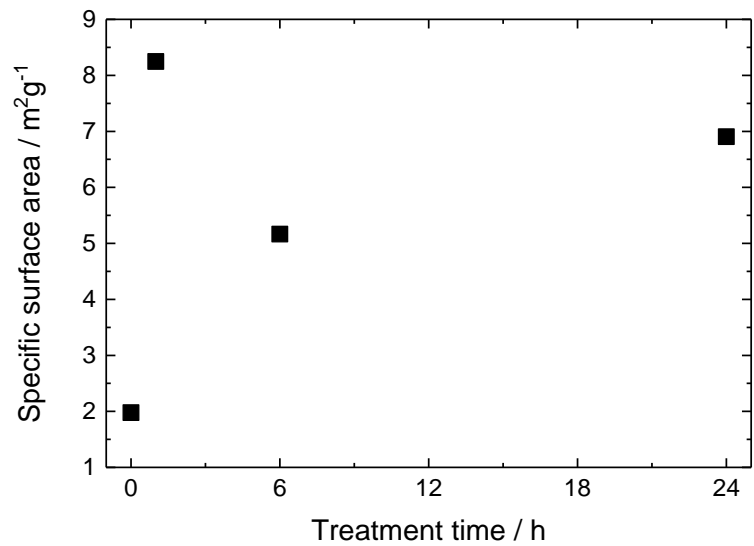

Figure 3. The specific surface area of raw and MC treated powders (MC-1-MC-24) calculated by BET methods from $N_2$ adsorption/desorption isotherms.

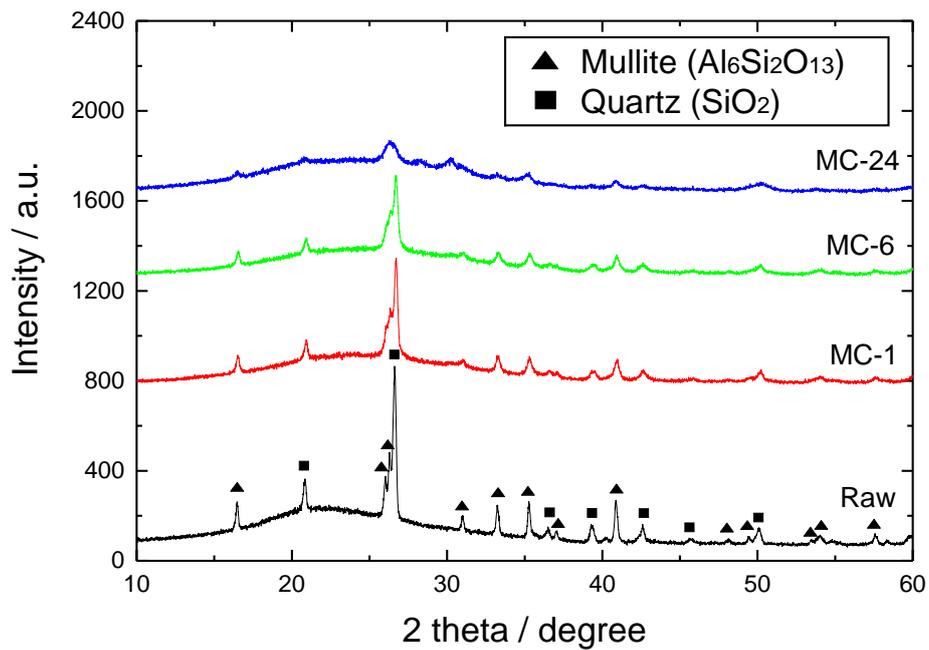

Figure 4. The XRD patterns of raw and MC treated powders (MC-1-MC-24).

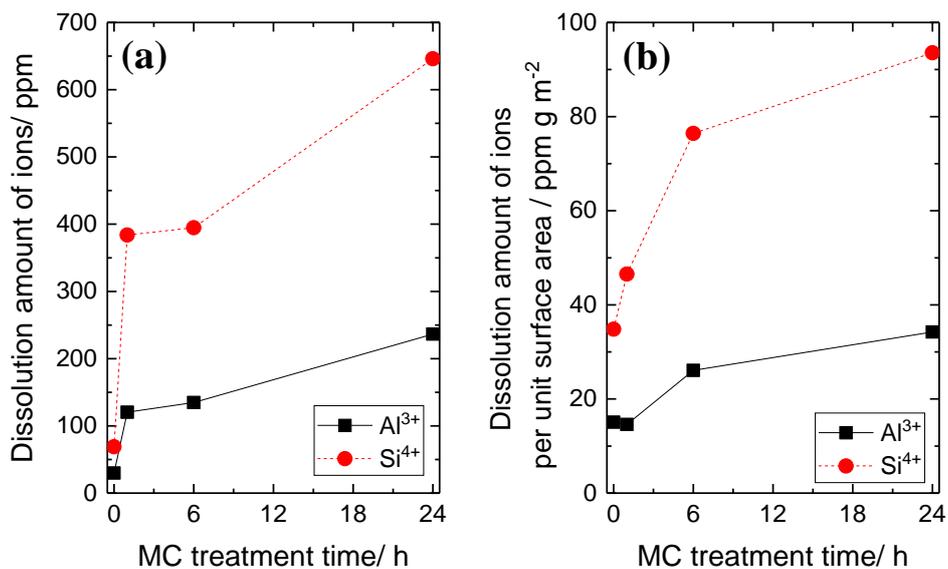

Figure 5. The dissolution amounts of ions into 3M NaOH solution (ICP-OES): (a) The dissolution amount of $Si^{4+}$ and $Al^{3+}$, (b) The dissolution amount of $Si^{4+}$ and $Al^{3+}$ per unit surface area.

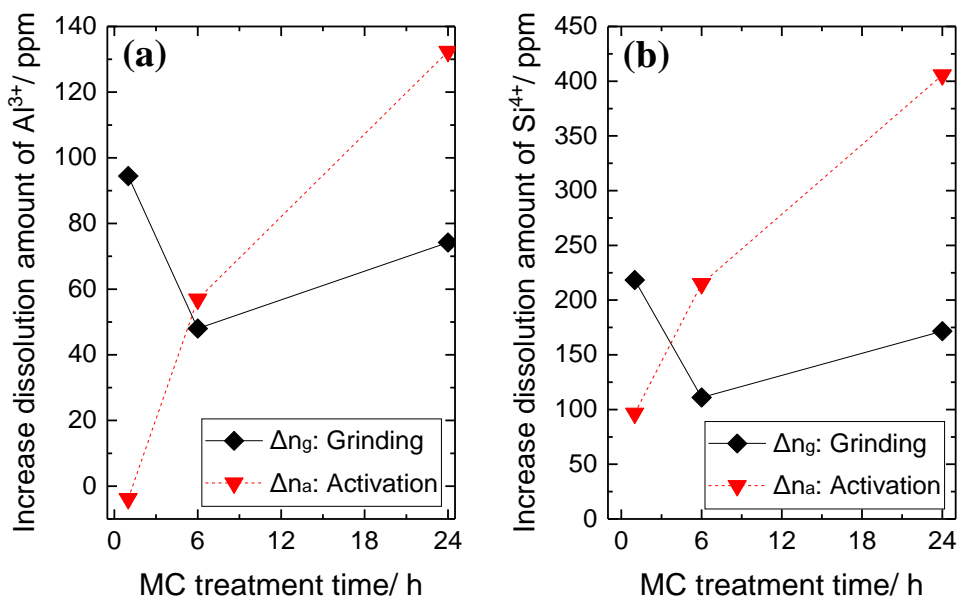

Figure 6. The dependence of dissolution amount on effects of milling and activation: The dissolution amount of (a) $Al^{3+}$ and (b) $Si^{4+}$.

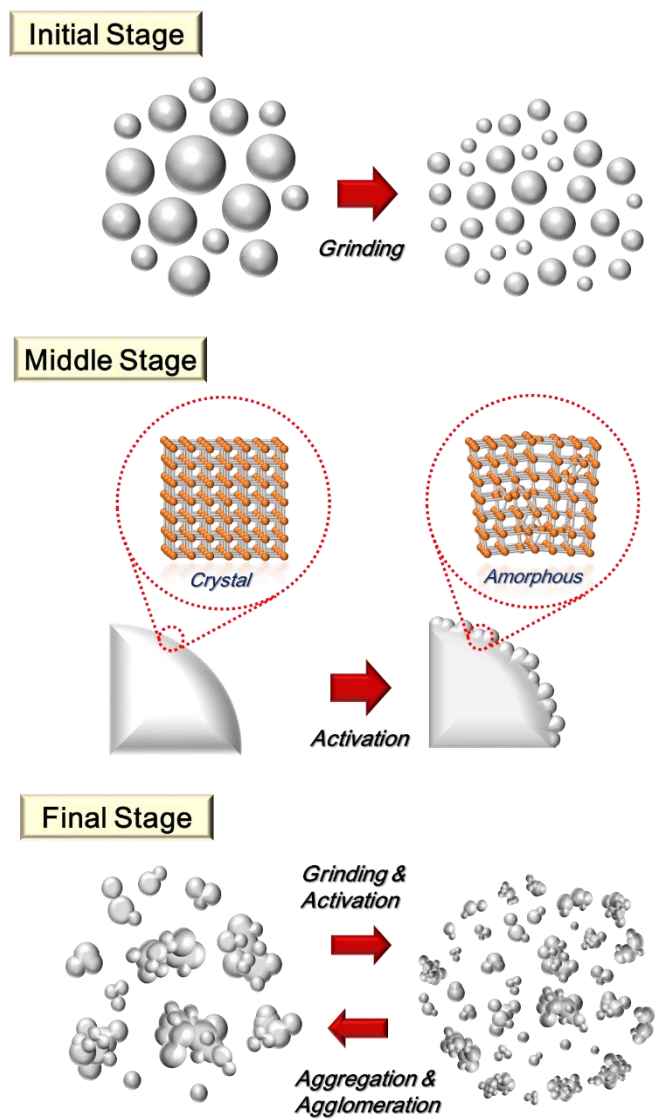

Figure 7. The schematic illustration of particle behavior during MC treatment.

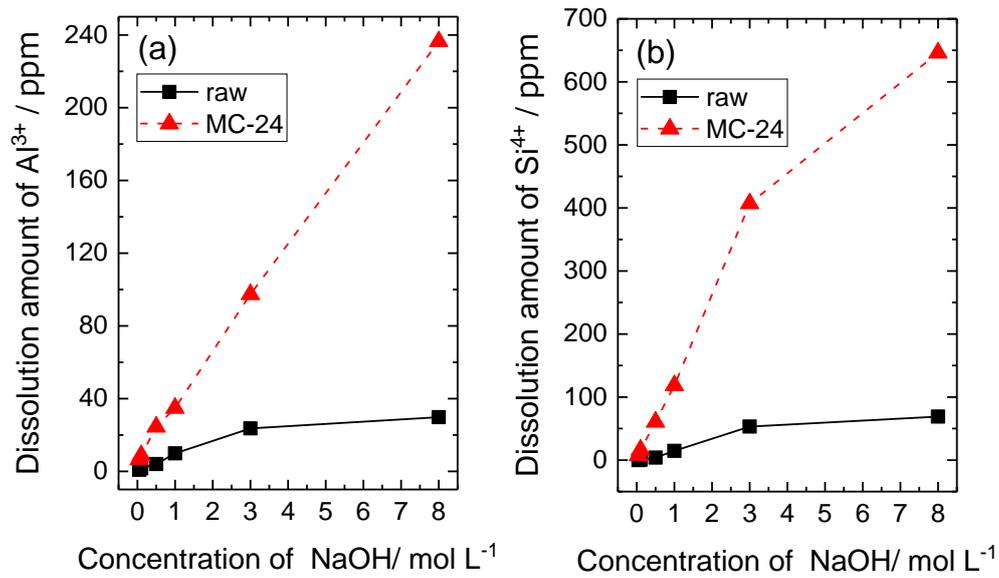

Figure 8. The dissolution amount with various concentration of NaOH solution: The dissolution amount of (a) $Al^{3+}$ and (b) $Si^{4+}$.